\documentclass{article}
\usepackage{knitr}
\usepackage{amsmath}
\usepackage{amssymb}
\usepackage{bm}
\usepackage{acronym}
\usepackage{booktabs}

\newcommand{\Cpp}{C\nolinebreak\hspace{-.05em}\raisebox{.4ex}{\small\bf +}\nolinebreak\hspace{-.10em}\raisebox{.4ex}{\small\bf +}}

\newcommand\mat[1]{\mathbf{#1}}
\renewcommand\vec{\bm}
\newcommand\der{\operatorname{d\!}{}}
\newcommand\bigO[1]{\mathcal{O}\left(#1\right)}

\DeclareMathOperator\Excp{E}

\DeclareMathOperator*{\argmax}{arg\,max}

\newcommand\nBreastObs{60081}

\acrodef      {MC}{Monte Carlo}

\acrodef      {KL}{Kullback–Leibler}

\acrodef      {VA}{variational approximation}
\acrodefplural{VA}{variational approximations}

\acrodef      {GVA}{Gaussian variational approximation}
\acrodefplural{GVA}{Gaussian variational approximations}

\acrodef      {BFGS}{Broyden-Fletcher-Goldfarb-Shanno}

\acrodef      {EM}{expectation maximization}

\acrodef      {MCMC}{Markov chain Monte Carlo}

\acrodef      {LMM}{linear mixed model}
\acrodefplural{LMM}{linear mixed models}

\acrodef      {GLMM}{generalized linear mixed model}
\acrodefplural{GLMM}{generalized linear mixed models}

\acrodef      {AGHQ}{adaptive Gauss–Hermite quadrature}

\IfFileExists{upquote.sty}{\usepackage{upquote}}{}

\usepackage{authblk}

\makeatletter
\newcommand{\manuallabel}[2]{\def\@currentlabel{#2}\label{#1}}
\makeatother

\usepackage{natbib}

\begin{document}

\title{
  Joint Models with Multiple Markers and Multiple Time-to-event Outcomes Using Variational Approximations
}

\date{\today}



\author[1,2]{Benjamin Christoffersen}
\author[1]{Keith Humphreys}
\author[1]{Alessandro Gasparini}
\author[1]{Birzhan Akynkozhayev}
\author[2]{Hedvig Kjellström}
\author[1]{Mark Clements}

\affil[1]{Department of Medical Epidemiology and Biostatistics, Karolinska Institutet}
\affil[2]{Division of Robotics, Perception and Learning, KTH Royal Institute of Technology}

\maketitle

\label{firstpage}


\begin{abstract}
Joint models are well suited to modelling linked data from laboratories and health registers. However, there are few examples of joint models that allow for (a) multiple markers, (b) multiple survival outcomes (including terminal events, competing events, and recurrent events), (c) delayed entry and (d) scalability. We propose a full likelihood approach for joint models based on a Gaussian variational approximation to satisfy criteria (a)-(d). We provide an open-source implementation for this approach, allowing for flexible sets of models for the longitudinal markers and survival outcomes. Through simulations, we find that the lower bound for the variational approximation is close to the full likelihood. We also find that our approach and implementation are fast and scalable. We provide an application with a joint model for longitudinal measurements of dense and fatty breast tissue and time to first breast cancer diagnosis. The use of variational approximations provides a promising approach for extending current joint models.
\end{abstract}


\section{Introduction}

Individual-based data linkages between health registers and laboratories are increasingly used in medical research, enabling joint modeling of longitudinal biomarkers and time-to-event outcomes. Joint models enhance efficiency and reduce bias, particularly when there is dependence between observation processes or terminal events and the biomarkers. Such dependence is important in cases of dropout, noisy measurements, or when the observation process is non-ignorable, irrespective of whether the primary interest lies in the time-to-event outcomes or the biomarkers.

Several software tools exist for fitting joint models, offering varying levels of flexibility and estimation methods. These tools support a broad range of models, including different longitudinal models (e.g., \acp{LMM} and \acp{GLMM}), survival models (e.g., non-parametric, parametric, or penalized baseline hazards, with covariates modeled via proportional hazards or accelerated failure time models), and statistical frameworks (e.g., frequentist or Bayesian). Estimation algorithms typically include full likelihood approaches for parametric models using maximum likelihood \citep{Rizopoulos10,garcia18} or Bayesian inference \citep{Rizopoulos16,Niekerk21}, as well as \ac{EM} algorithms for non-parametric baseline hazards \citep{Rizopoulos10,kim17,Hickey18}.

Existing tools effectively handle many model configurations, yet certain challenges persist, particularly with scalability for larger datasets or high-dimensional models involving multiple longitudinal markers and outcomes. Bayesian approaches, such as JMBayes2 \citep{Rizopoulos24} and integrated nested Laplace approximations in INLA \citep{Rustand23}, offer advanced capabilities for complex scenarios, including competing risks and non-Gaussian data. However, \ac{MCMC} based methods may require substantial computation time for large datasets with complex models. In the frequentist framework, flexible options remain limited. Recent advancements, such as the \ac{VA}-based method introduced by \cite{Tu23}, provide a modular and efficient approximation approach in the frequentist setting.

This paper investigates an approximate likelihood estimation method aimed at addressing specific challenges in a frequentist setting: \emph{(a) modeling multiple longitudinal markers; (b) incorporating multiple survival outcomes; (c) handling delayed entry; and (d) ensuring scalability for larger datasets}. Pror studies have developed sophisticated methods addressing individual aspects of these challenges. For (a), multivariate extensions of joint models for longitudinal and time-to-event outcomes have been reviewed by \citet{Hickey16}, with implementations such as \citet{Hickey18}. For parametric survival models, a variety of models have been developed for multiple longitudinal markers \citep{Rizopoulos16,Krol2017,brilleman19,crowther20}. For (b), examples of joint models with multiple survival outcomes include recurrent and terminal events or multi-state data \citep{Krol2017,Ferrer16,crowther20}. For (c), methods accommodating delayed entry are described in \citep{Rizopoulos16,Crowther16,proust17,crowther20}. For (d), the scalability of Bayesian and \ac{EM}-based approaches \citep[see][for an approximation]{MURRAY22} remains a limitation, particularly for high-dimensional random effects.

We investigate the use of \ac{VA} \citep{Ormerod10} for joint modeling of longitudinal Gaussian outcomes to assess its potential to achieve the desired goals. \ac{GVA} transforms the likelihood maximization problem involving many intractable integrals into a high-dimensional optimization problem that can be efficiently evaluated pointwise. This approach has been explored by \cite{Tu23} using a coordinate ascent algorithm commonly applied in \ac{VA} \citep{Ormerod12}.

We extend the methodology by providing a package that supports a broader range of models and incorporates a tailored quasi-Newton method for jointly optimizing all parameters. The resulting approach offers significant speed improvements and demonstrates applicability of \ac{VA} to clustered models.

The remainder of this paper is organized as follows. Section~\ref{sec:model} describes the longitudinal and survival sub-models and the likelihood. Section~\ref{sec:VA} introduces the use of \ac{GVA} for joint models, detailing inference and how to estimate them fast. Section~\ref{sec:simNApplication} evaluates the small-sample performance of our approach through simulations and applies it to a well-known benchmarking dataset and a large breast cancer of longitudinal markers from mammographic images. We conclude with a discussion of the method and potential extensions in Section~\ref{sec:discussion}.


\section{Model}\label{sec:model}

The sub-models for the longitudinal variables
and survival outcomes are introduced in this
section. The models are multivariate extensions of both
the longitudinal variable sub-models
and the survival sub-models presented by
\citet{Henderson00}. The longitudinal sub-model is a multivariate
\ac{LMM}, while the survival sub-model is a multivariate flexible
parametric proportional hazards model.
The intractable log marginal likelihood for the joint model,
which is used to motivate our suggested \ac{VA} method, is presented
in Section \ref{subsec:MargL}.

\subsection{Longitudinal variable sub-models}\label{subsec:LMM}

Let $i$ be an index for the observed individuals, where there are $n$
individuals. Each individual have $L$ longitudinal measures at
$m_i$ times, with observation times $s_{i1},\dots,s_{im_i}$. For now,
all longitudinal variables are measured at each time $s_{ij}$ to keep the
notation simple and we later discuss the situation where only a subset
are measured. Let $l=1,\dots,L$ be an
index for the longitudinal variables. We assume that each variable has an
unobserved random effect $\vec U_{il}\in\mathbb R^{r_l}$, where $r_l$ is the
dimension of the random effect.
The longitudinal variable $l$ of individual $i$ at
time $s_{ij}$ is denoted by $Y_{ijl}$, which we assume is
normally distributed given the random effect $\vec U_{il} = \vec u_{il}$,
such that
$
\left(Y_{ijl} \mid \vec U_{il} = \vec u_{il}\right)
  = \mu_{il}(s_{ij}, \vec u_{il}) + \epsilon_{ijl},
$
where
$\mu_{il}(s, \vec u) = \vec x_{il}(s)^\top\vec\beta_l + \vec m_{il}(s)^\top\vec u$
and $\epsilon_{ijl} \sim N(0, \sigma_l^2)$.
$\mu_{il}(s, \vec u)$ is the latent mean curve for individual $i$ and
longitudinal variable $l$,
with linear predictor
$\vec x_{il}(s)^\top\vec\beta_l$ for the fixed effects, and linear predictor
$\vec m_{il}(s)^\top\vec U_{il}$ for the random effect.
$N(0, \sigma_l^2)$ denotes a normal distribution with mean $0$ and
variance $\sigma_l^2$.

One example of a parameterization of the fixed effects is
$\vec x_{il}(s) = (\tilde{\vec x}_{il}^\top,\allowbreak \vec g_l(s)^\top)^\top$
where
$\tilde{\vec x}_{il}^\top$ contains individual-specific covariates and
$\vec g_l$ is a given expansion (e.g., a polynomial such as $\vec{g_l(s)}=(1,s)^\top$ or a spline basis)
to allow for a time-varying intercept.
An alternative is
$\vec x_{il}(s) = (\tilde{\vec x}_{il}^\top, \vec g_l(s)^\top,\allowbreak f_{il}\vec g_l^f(s)^\top)^\top$
for some covariate $f_{il}$ and expansion $\vec g_l^f(s)$ leading to a so-called
"varying-coefficient" for $f_{il}$.
$\vec m_{il}(s)$ can be defined similarly to allow for
time-constant or time-varying deviations from the population.
For example, $\vec m_{il}(s) = (1, s)$ yields a random intercept and slope.

To define the joint model of all $L$ longitudinal variables, we stack the values, such that
$\vec U_i = (\vec U_{i1}^\top,\allowbreak \dots,\allowbreak \vec U_{iL}^\top)^\top\allowbreak \in \mathbb R^R$
with $R = \sum_{l = 1}^L r_l$. Similarly, we stack the fixed effects, the
observed values and error terms to get $\vec\beta$, $\vec Y_{ij}$, and
$\vec\epsilon_{ij}$, respectively. Furthermore, the design matrix
for individual $i$ for the fixed effects is %
$$
\mat X_i(s) = \begin{pmatrix}
  \vec x_{i1}(s)^ \top & \vec 0^\top & \cdots & \vec 0^\top \\
  \vec 0^\top & \vec x_{i2}(s)^\top & \ddots & \vdots \\
  \vdots & \ddots & \ddots & \vec 0^\top \\
  \vec 0^\top & \cdots & \vec 0^\top & \vec x_{iL}(s)^\top
  \end{pmatrix}.
$$%
The design matrix for the random effects, $\mat M_i(s)$, has a similar
construction. Let the covariance matrices for $\vec\epsilon_{ij}$ and
$\vec U_i$ be $\vec\Sigma$ and $\vec\Psi$, respectively.
Then the conditional joint distribution of $\vec Y_{ij}$ at
time $s_{ij}$ given the random effects $\vec U_i = \vec u_i$ is%
\begin{align*}
\left(\vec Y_{ij} \mid \vec U_i = \vec u_i \right)
  &= \vec\mu_i(s_{ij}, \vec u_i) + \vec\epsilon_{ij}
  & \vec\epsilon_{ij} &\sim N^{(L)}(\vec 0, \mat\Sigma) \\
\vec\mu_i(s_{ij}, \vec u_i) &= \mat X_i(s_{ij})\vec\beta +
  \mat M_i(s_{ij})\vec u_i & \vec U_i &\sim N^{(R)}(\vec 0, \mat\Psi)
\end{align*}%
where $N^{(Q)}(\vec 0, \mat\Psi)$ denotes the $Q$-dimensional normal
distribution
with mean vector $\vec 0$ and covariance matrix $\mat\Psi$.
The errors, $\vec\epsilon_{ij}$, may be correlated making the longitudinal
variables observed at the same time point marginally dependent.
Moreover, the variables may be
marginally dependent across time through $\mat\Psi$.
A block diagonal structure can be assumed for $\mat\Psi$ making only variables
of the same type marginally dependent across time.
It is common to
observe only some of the $L$ longitudinal variables,
and often only one of the $L$ longitudinal variables, at
each time $s_{ij}$.
Assuming that the missingness is not related to the value of $Y_{ijl}$,
the conditional density shown in Section
\ref{subsec:MargL} can easily  be adjusted
because the marginal distribution is also normally distributed.
However, the covariances of $\mat\Sigma$ are not identifiable if two pairs
of variables, $Y_{ijl}$ and $Y_{ijl'}$,
are never observed at the same point in time.

\subsection{Survival sub-models}

We describe the parameterizations of the survival models we use. A motivation
for the parameterizations compared to other common parameterizations
is provided in Appendix \ref{sec:alternativeParam}.
We allow for $E$ different time-to-event outcomes. Each type
can be terminal, competing or recurrent. We define the model for all types in
terms of their conditional hazard.
The conditional hazard for the $e$-th outcome for individual $i$ given the
random effects
$\vec U_i = \vec u_i$ and possibly a frailty $\exp W_{ie} = \exp w_{ie}$
(see Section \ref{subsec:recurrent} and \ref{subsec:frailty})
is of the form %
\begin{equation}
h_{ie}(t\mid \vec u_i, w_{ie}) = \exp\left(
  \vec z_{ie}(t)^\top\vec \gamma_e +
  \vec\alpha_e^\top \mat M_i(t)\vec u_i + w_{ie}
  \right) \label{eqn:subhaz}
\end{equation}%
where $W_{ie} \sim N(0,\xi_e^2)$ is a random effect,
$\vec z_{ie}(t)$ and $\vec\gamma_e$ are the design vector and coefficients
for the fixed effects, respectively, and
$\vec\alpha_e$ and $\mat M_i(t)$ are the
association parameters and design matrix for the shared random effects defined
for the multivariate \ac{LMM} in Section \ref{subsec:LMM}, respectively.
The design vector $\vec z_{ie}(t)$ may have a similar parameterization to
$\vec x_{il}(t)$. The form
$\vec z_{ie}(t) = (\tilde{\vec z}_{ie}^\top, \vec b_e(t)^\top)^\top$ is
common with covariates $\tilde{\vec z}_{ie}$ which have a conditional
proportional hazard and an expansion
$\vec b_e(t)$ for the log baseline
hazard. This is the form we use in this paper where $\vec b_e(t)$ is chosen to be
a B-spline and $\tilde{\vec z}_{ie}$ are covariates with a conditional
proportional hazard.
The design vector $\vec z_{ie}(t)$ can, as for $\vec x_{il}(t)$, be
extended to
have varying-coefficients for some of the covariates
(i.e.\ non-proportional hazards). This is currently not
supported by our software but we are working on this extensions.
The
association parameters, $\vec\alpha_e$, link the deviations from the
population mean of the longitudinal variables
to the log-hazards of a given individual.

The formula in Equation~\eqref{eqn:subhaz} links the hazard to the current deviation. However, research may focus on the cumulative value, $(\int_0^t \mat M_i(s) \der s)\vec U_i$, the derivative, $(\partial \mat M_i(t) \big/\partial t)\vec U_i$, or both, depending on specific hypotheses. This choice depends on the application and prior knowledge. In some cases, instantaneous risk may be tied to marker changes, favoring the derivative. In others, long-term exposure to elevated markers may increase instantaneous risk, making the cumulative value more relevant.

As an example of how this looks in practice, suppose that
there are $L = 2$ longitudinal variables and
we assume that survival type $e$
has a hazard that
depends on the current value and slope for the first longitudinal variable,
and depends on the cumulative value for the
second variable. Then we replace $\mat M_i(t)$
in Equation \eqref{eqn:subhaz} with $\mat B_{ie}(t)$ defined by %
$$
\mat B_{ie}(t) = \begin{pmatrix}
  \vec m_{i1}(t)^ \top & \vec 0^\top  \\
  \left.\partial\vec m_{i1}(s)^\top \big/ \partial s \right\vert_{s = t}
    & \vec 0^\top  \\
  \vec 0^\top & \int_0^t \vec m_{i2}(s)^ \top \der s  \\
  \end{pmatrix}.
$$%
This adds one extra association parameter to
$\vec\alpha_e = (\alpha_{e11}, \alpha_{e12}, \alpha_{e21})^\top$.

\subsubsection{Terminal and competing events}

When event type $e$ is terminal for individual $i$, we
let $T_{ie}^*$ be the true time-to-event outcome, for type $e$, and
$C_i$ be the right-censoring time of individual $i$.
We assume that $T_{ie}^*$ and $C_i$ are independent and
let  $T_{ie} = \min(T_{ie}^*, C_i)$ be the observed time and
$\delta_{ie} = 1_{\{T_{ie}^* \leq  C_i\}}$ be an event indicator.
The log of the conditional density of the outcome is then %
\begin{equation}\label{eqn:CompleteLLSurv}
b_{T_{ie}}(t_{ie}, \delta_{ie}, \vec u_i, w_{ie}) =
  \delta_{ie}\log h_{ie}(t_{ie}\mid \vec u_i, w_{ie}) -
  \int_{0}^{t_{ie}}h_{ie}(s\mid \vec u_i, w_{ie}) \der s.
\end{equation}%
The integral for the cumulative hazard in Equation \eqref{eqn:CompleteLLSurv}
is intractable for all models we use but can be efficiently approximated with
quadrature rules such as
Gauss–Legendre or Gauss–Kronrod \citep{Crowther14}. The
former is used in this paper.

Our framework can incorporate more than one type of terminal event,
which is necessary for modelling cause-specific
deaths and competing risks in general.
If types $e \in \mathcal C \subseteq \{1,\dots,E\}$ are competing, we
assume that the events are conditionally independent and define
$\bar C_i = \min(C_i, \allowbreak \min_{e \in \mathcal C} \allowbreak T_{ie}^*)$
and redefine
the event indicator as $\delta_{ie} = 1_{\{T_{ie}^* \leq  \bar C_i\}}$. Then
Equation \eqref{eqn:CompleteLLSurv} still represents the
log conditional density of the outcome when summed
over the $e\in \mathcal C$ terms.

\subsubsection{Recurrent events}\label{subsec:recurrent}
An example of a recurrent event that one
may want to model jointly with marker values is the
observation process.
In observational
studies, the observation process may be correlated with the
severity of the disease, through mutual dependence on the
longitudinal variables.
The joint model can be used to account for such dependencies
\citep[see][]{Gasparini20}. If type $e$ is
a recurrent event,
we let $T_{iej}^*$ be the event time $j$ of type $e$ for individual
$i$,  $k_{ie}$ be the number of observed recurrent events before $\bar C_i$,
$T_{iej} = \min (T_{ihj}^*, \bar C_i)$ be the observed time and
$\delta_{iej} = 1_{\{T_{iej}^* \leq  \bar C_i\}}$ be the event indicator for
event $j$.
Let $t_{ie0} = 0$ by definition.
Then the log of the conditional density of the outcome is
\begin{multline}\label{eqn:recurrentLL}
b_{T_{ie}}(t_{ie1}, \cdots, t_{iek_{ie}},
  \delta_{ie1}, \cdots, \delta_{iek_{ie}} \vec u_i, w_{ih}) \\
= \sum_{j = 1}^{k_{ie} + 1}
  \delta_{iej}\log h_{ie}(t_{iej}\mid \vec u_i, w_{ie}) -
  \int_{t_{ie,j-1}}^{t_{ieh}}h_{ie}(s\mid \vec u_i, w_{ie}) \der s.
\end{multline}

\subsubsection{Frailties}\label{subsec:frailty}

The random effects $\vec W_i = (W_{i1}, \dots, W_{iE})^\top$
are assumed to follow a normal distribution
with a zero mean vector and covariance matrix $\mat\Xi\in\mathbb R^{E\times E}$.
Their purpose is to allow for random effects with recurrent events and to
make the time-to-event outcomes marginally
dependent even if the association parameters are all zero.
In principle, the
random effects could also be used with terminal outcomes in which case they
would
provide evidence of non-proportional hazards and heterogeneity. However,
this would require there to be
variation in the hazards due to individual-specific
covariates \citep{Elbers82}.
Moreover, the covariances between two random variables $W_{ie}$ and $W_{ie'}$
would not be identifiable if survival type $e$ and $e'$  ($e\neq e'$)
are competing risks \citep{Crowder91}.
These covariances will, of course, not be included.

\subsection{Marginal likelihood}\label{subsec:MargL}
The log marginal likelihood can
be written as
\begin{equation}\label{eqn:margLNoDelay}
\bar L_i(\vec\zeta) = \log\Excp_{p_{i, \vec\zeta}}\Big(
  \exp\big(
  b_{\vec T_i}(\vec t_i, \vec\delta_i, \vec U_i, \vec W_i)
  + b_{\mat Y_i}(\mat y_i, \vec U_i)
  \big)\Big)
\end{equation}%
where $\vec\zeta$ is the vector of model parameters, $p_{i, \vec\zeta}$ denotes the true distribution under $\vec\zeta$, and the subscript in the expectation specifies the distribution used, integrating out the random effects.
$b_{\vec T_i}$ is the log of the conditional density of the survival outcomes
and $b_{\mat Y_i}$ is the log of the conditional density of the longitudinal
outcomes. Both densities are covered below.

The log of the conditional density for all time-to-event outcomes of individual
$i$ is%
$$
b_{\vec T_i}(\vec t_i, \vec\delta_i, \vec u_i, \vec w_i) =
  \sum_{e = 1}^{E}
  \begin{cases}
    b_{T_{ie}}(t_{ie}, \delta_{ie}, \vec u_i, w_{ie}) & e \in \mathcal C \\
    b_{T_{ie}}(t_{ie1}, \cdots, t_{iek_{ie}},
    \delta_{ie1}, \cdots, \delta_{iek_{ie}} \vec u_i, w_{ie}) &
      \text{otherwise}
  \end{cases}
$$%
where the log densities are defined in Equation \eqref{eqn:CompleteLLSurv} and
\eqref{eqn:recurrentLL}.

Let $\phi^{(Q)}(\vec x;\vec\xi,\mat\Omega)$ be the $Q$ dimensional
normal distribution's density function
with mean $\vec\xi$ and covariance matrix $\mat\Omega$ evaluated at
$\vec x$. Then the log of the conditional density
for observation $j$ of the longitudinal variables of individual $i$ is
$b_{\vec Y_{ij}}(\vec y_{ij}, \vec u_i) =
  \log\phi^{(L)}\left(\vec y_{ij}; \vec\mu_i(s_{ij}, \vec u_i), \mat\Sigma\right)$
and the contribution of all the $m_i$ observations is %
\begin{equation}\label{eqn:CondMarker}
b_{\mat Y_i}(\mat y_i, \vec u_i) = \sum_{j = 1}^{m_i}
  b_{\vec Y_{ij}}(\vec y_{ij}, \vec u_i).
\end{equation}%
The marginal likelihood is intractable because of the survival part of the
model.

\subsubsection{Delayed entry}

Delayed entry is common in observational studies
(see the example in Section \ref{sec:application},
using attained age as the time scale).
Let $v_i\geq 0$ be the delayed entry time for
individual $i$. Then the log marginal likelihood differs from Equation
\eqref{eqn:margLNoDelay} \citep{Crowther16} and is given by %
\begin{align}
L_i(\vec\zeta) &= \bar L_i(\vec\zeta) - D_i(\vec\zeta)\label{eqn:margL} \\
D_i(\vec\zeta) &= \log\Excp_{p_{i, \vec\zeta}}\Big(
  \exp(b_{V_i}(v_i, \vec U_i, \vec W_i))
  \Big)\label{eqn:delayedTerm}
\end{align}%
where
$
b_{V_i}(v_i, \vec U_i, \vec W_i) =
  -\sum_{e \in \mathcal C}
    \int_{0}^{v_i}h_{ie}(s\mid \vec u_i, w_{ie}) \der s
$.
The conditional density for the recurrent events also needs to be changed to
$t_{ie0} = v_i$. Moreover, $T_{ie1}$ is the observed time for the first
recurrent event after $v_i$ if $e\in\{1,\dots,E\}\setminus\mathcal C$.
Thus, two intractable integrals have to be approximated when there
is delayed entry.


\section{Variational approximations}\label{sec:VA}

We suggest to approximate the log marginal likelihood in Equation
\eqref{eqn:margL} using \acp{VA}. Following \cite{Ormerod10},
let $\vec O_i = (\vec U_i^\top, \vec W_i^\top)^\top$ be the
concatenated vector of random effects for individual $i$,
$\exp L_i(\vec\zeta) = p_{i, \vec\zeta}(\mat y_i, \vec t_i)$ be the marginal
likelihood factor of individual $i$ evaluated at $\vec\zeta$, and
$p_{i, \vec\zeta}(\mat y_i, \vec t_i, \vec o_i)$ be the complete
data likelihood given by %
\begin{equation}\label{eqn:CompleteData}
p_{i, \vec\zeta}(\mat y_i, \vec t_i, \vec o_i) =
  \exp\Big(
    b_{\vec T_i}(\vec t_i, \vec\delta_i, \vec U_i, \vec W_i)
    + b_{\mat Y_i}(\mat y_i, \vec U_i) +
    b_{\vec U_i,\vec W_i}(\vec u_i, \vec w_i)
    \Big)
\end{equation}
where
\begin{equation}\label{eqn:logUncond}
b_{\vec U_i,\vec W_i}(\vec u_i, \vec w_i) =
  \log\phi^{(R)}(\vec u_i; \vec 0, \mat\Psi) +
  \log\phi^{(H)}(\vec w_i;\vec 0, \mat\Xi).
\end{equation}%
The version with delayed entry is given in Section
\ref{subsec:delayedEntry}.
Let $p_{i, \vec\zeta}(\vec o_i \mid \mat y_i, \vec t_i)$ be the
conditional
density of the random effects given the observed outcomes and
$q_{\vec\theta_i}(\vec o_i)$ be the density of a chosen variational distribution
indexed by $\vec\theta_i\in\Theta$ which has support on $\mathbb R^{R + E}$.
Then %
\begin{align}
L_i(\vec\zeta) &= \int q_{\vec\theta_i}(\vec o)
  \log\left(\frac
  {p_{i, \vec\zeta}(\mat y_i, \vec t_i, \vec o)\big/q_{\vec\theta_i}(\vec o)}
  {p_{i, \vec\zeta}(\vec o \mid \mat y_i, \vec t_i)\big/q_{\vec\theta_i}(\vec o)}
  \right)\der\vec o
  \nonumber\\
&= \int q_{\vec\theta_i}(\vec o)
  \log\left(\frac
    {p_{i, \vec\zeta}(\mat y_i, \vec t_i, \vec o)}
    {q_{\vec\theta_i}(\vec o)}
  \right)\der\vec o +
  \int q_{\vec\theta_i}(\vec o)
  \log\left(\frac
    {q_{\vec\theta_i}(\vec o)}
    {p_{i, \vec\zeta}(\vec o \mid \mat y_i, \vec t_i)}
  \right)\der\vec o
  \nonumber\\
&\geq \int q_{\vec\theta_i}(\vec o)
  \log\left(\frac
    {p_{i, \vec\zeta}(\mat y_i, \vec t_i, \vec o)}
    {q_{\vec\theta_i}(\vec o)}
  \right)\der\vec o = \tilde{L}_i(\vec\theta_i,\vec\zeta).
  \label{eqn:LBBroad}
\end{align}%
The inequality follows as the term we removed is the \ac{KL}
divergence between the conditional density for the random effects,
$p_{i, \vec\zeta}(\vec o_i \mid \mat y_i, \vec t_i)$,
and the density of the variational distribution,
which is greater than or equal to zero. Thus,
there is equality
if the conditional distribution of the random effects is identical to the
variational distribution.

As we show here, the lower bound is computationally very easy to evaluate
pointwise for some variational distributions.
Thus, one may quickly obtain an approximation of the maximum likelihood
estimator from $n$ individuals with %
\begin{equation}\label{eqn:maxLBEst}
\argmax_{\vec\zeta} \sum_{i = 1}^{n} \max_{\vec\theta_i}
  \tilde{L}_i(\vec\theta_i,\vec\zeta).
\end{equation}%
The optimization problem easily becomes high dimensional when seen as a
joint
optimization problem over $\vec\zeta,\vec\theta_1,\dots,\vec\theta_n$. However,
the optimization problem is so-called
partially separable and this can be used to quickly
estimate the model
using tailored quasi-Newton methods \citep{nocedal06}.
We use the psqn package \citep{Christoffersen21}
we have developed which provides
an implementation of such a quasi-Newton method.
Further details of the quasi-Newton method
are provided in Appendix \ref{sec:psqnNotes}.
Our implementation is in \Cpp~with a small R interface. The implementation
supports computation in parallel using OpenMP. We use an extension of the
reverse mode automatic differentiation library by \cite{savine2018modern}
to get the gradient.
See Appendix \ref{sec:implementation} for further details.

\subsection{Lower bound terms}
There are three types of lower bound terms from Equation \eqref{eqn:LBBroad}.
The overall lower bound is%
\begin{equation}\label{eqn:LBtermsFull}
\tilde{L}_i(\vec\theta_i,\vec\zeta) =
  \Excp_{q_{\vec\theta_i}}\Big(
  b_{\vec T_i}(\vec t_i, \vec\delta_i, \vec U_i, \vec W_i)
  + b_{\mat Y_i}(\mat y_i, \vec U_i) +
b_{\vec U_i,\vec W_i}(\vec u_i, \vec w_i)
  - \log q_{\vec\theta_i}(\vec O_i)
  \Big).
\end{equation}%
The first type of term is the entropy of the
variational distribution,
$-\Excp_{q_{\vec\theta_i}}\left(\log q_{\vec\theta_i}(\vec O_i)\right)$.
The second type is from the longitudinal variables and
the unconditional distribution of the random effects from Equation
\eqref{eqn:CondMarker} and \eqref{eqn:logUncond}.
Their terms in the lower bound are given by %
\begin{multline*}
\Excp_{q_{\vec\theta_i}}\left(
  \log\phi^{(Q)}(\vec x;\vec\xi + \mat S\vec O_i,\mat\Omega)\right) = \\
  -\frac 12 \left(
  Q\log(2\pi) + \log\lvert\mat\Omega\rvert +
  \Excp_{q_{\vec\theta_i}}\left(
    -(\vec x - \vec\xi - \mat S\vec O_i)^\top\mat\Omega^{-1}
    (\vec x - \vec\xi - \mat S\vec O_i)
  \right)\right)
\end{multline*} %
for given matrices $\mat S$, $\mat\Omega$ and vector
$\vec\xi$. This requires evaluations of the first and second moments.
Lastly, the hazard in Equation \eqref{eqn:subhaz} can be written as
$h_{ie}(t\mid\vec o_i) = \exp\left(g_{ie}(t) + \vec a_{ie}(t)\vec o_i\right)$
for suitably chosen $g_{ie}(t)$ and $\vec a_{ie}(t)$ which implicitly
depend on the parameters in the model. Thus, the expectation of
Equation \eqref{eqn:CompleteLLSurv} and \eqref{eqn:recurrentLL} reduces to
sums over terms of the form%
\begin{multline} \label{eqn:expecHaz}
\Excp_{q_{\vec\theta_i}}\left(
  d\log h_{ie}(t\mid \vec O_i) -
  \int_c^th_{ie}(s\mid \vec O_i) \der s
  \right) = \\
d\left(g_{ie}(t) + \vec a_{ie}(t)\Excp_{q_{\vec\theta_i}}(\vec O_i)\right)
  -\int_c^t
  \Excp_{q_{\vec\theta_i}}\left(
  \exp\left(g_{ie}(s) + \vec a_{ie}(s)\vec O_i
  \right)\right)
  \der s
\end{multline}%
for a given event indicator $d$, left-truncation time $c$, and event time $t$ and
assuming that we can interchange the order of integration. Thus, we need
the first moment and the moment generating function of the variational
distribution. Importantly, we only have
to do numerical integration in one dimension if the moment generating function
of the variational distribution is easy to evaluate. The
remaining integration for the expected cumulative hazard can easily
be solved with a quadrature rule as in the conditional case in Equation
\eqref{eqn:CompleteLLSurv}.

We use multivariate normal distribution with
$q_{\vec\lambda_i, \mat\Lambda_i}(\vec o_i) =
  \phi^{(R + E)}(\vec o_i;\vec\lambda_i,\mat\Lambda_i)$ and call it a
\ac{GVA} as per \cite{Ormerod12} and others. The  multivariate normal
distribution has a lower bound
which is very fast to evaluate pointwise because it has a closed-form solution
for the entropy, first two moments, and moment generating
function. Moreover, as we will show later, it yields an almost exact maximum likelihood estimator in
many cases. This is perhaps not too surprising given that the
conditional distribution of the random effects is a multivariate normal
distribution in the longitudinal variable only model.

\subsection{Inference}
The lower bound in Equation \eqref{eqn:maxLBEst} can be used to
quickly construct
an approximation of the observed information which
can be used to construct Wald tests and confidence intervals.
To see this, we can view the
optimization problem in Equation
\eqref{eqn:maxLBEst} as
$\argmax_{\vec\zeta} \sum_{i = 1}^n
  \hat{L}_i(\vec\zeta)$ with
$\hat{L}_i(\vec\zeta) =
  \tilde L_i\left(\hat{\vec\theta}_i(\vec\zeta), \vec\zeta\right)$
where
$\hat{\vec\theta}_i(\vec\zeta) = \argmax_{\vec\theta} \tilde L_i\left(\vec\theta, \vec\zeta\right)$.
This can be used to show that the
Hessian of $\hat{L}_i(\vec\zeta)$ is%
\begin{equation}\label{eqn:obsMat}
\left.
    \nabla^2_{\vec\zeta\vec\zeta}
      \tilde L_i\left(\vec\theta, \vec\zeta\right)
    -\nabla^2_{\vec\zeta\vec\theta}
     \tilde L_i\left(\vec\theta, \vec\zeta\right)
     \left(\nabla^2_{\vec\theta\vec\theta}
           \tilde L_i\left(\vec\theta, \vec\zeta\right)
     \right)^{-1}
     \nabla^2_{\vec\theta\vec\zeta}
     \tilde L_i\left(\vec\theta, \vec\zeta\right)
  \right\rvert_
  {\vec\theta = \hat{\vec\theta}_i(\vec\zeta)}
\end{equation}%
using that
$\nabla_{\vec\theta}
  \tilde L_i\left(\vec\theta, \vec\zeta\right)\rvert_
  {\vec\theta = \hat{\vec\theta}_i(\vec\zeta)} =\vec 0$. This is
computationally straightforward to compute and can be computed with numerical
differentiation given a gradient implementation as $\tilde L_i$ is not
high-dimensional.
Our simulation studies in Section \ref{sec:Sims}
show that the observed information matrix in
Equation \eqref{eqn:obsMat} is
very close to that obtained from a more precise but much slower
quadrature-based implementation.
The lower bound can also be used to construct
approximate profile likelihood based confidence intervals. In our simulation
studies in Section \ref{sec:Sims}, we find no evidence of such confidence
intervals not having nominal coverage for the association parameters.

Another benefit of \acp{VA} is that the variational parameters
$\hat{\vec\theta}_i(\vec\zeta)$ are
the $\vec\theta_i\in\Theta$ that
minimize the \ac{KL} divergence between the conditional distribution
of the random effects
and the variational distribution.
Thus, the maximum likelihood estimator of the
model parameters, $\vec\zeta$, and the variational parameters,
$\hat{\vec\theta}_i(\vec\zeta)$, can be used to compute approximate
marginal measures. An example is
the conditional distribution of the mean curve of the longitudinal variables.
Such computations can be done extremely fast
as the variational distribution we use is multivariate normal.

\subsubsection{Delayed entry}\label{subsec:delayedEntry}
The complete data likelihood shown in Equation \eqref{eqn:CompleteData} is
changed to %
$$
p_{i, \vec\zeta}(\mat y_i, \vec t_i, \vec o_i) =
  \frac{\exp\Big(
    b_{\vec T_i}(\vec t_i, \vec\delta_i, \vec U_i, \vec W_i)
    + b_{\mat Y_i}(\mat y_i, \vec U_i) +
    b_{\vec U_i,\vec W_i}(\vec u_i, \vec w_i)
    \Big)}
  {\exp(D_i(\vec\zeta))}
$$%
when there is delayed entry. Thus, the lower bound in Equation
\eqref{eqn:LBtermsFull} has an additional $-D_i(\vec\zeta)$ term and,
therefore,
we also need an approximation of the log of the expectation in Equation
\eqref{eqn:delayedTerm}. A \ac{VA}
can also be used to approximate the additional terms.
The terms of the lower bound on $D_i(\vec\zeta)$
are like those presented in the previous section and it
introduces new variational parameters for each individual which we denote by
$\vec\rho_1,\dots,\vec\rho_n$. However, this changes
the optimization problem. Before we exploit that we can write the optimization
problem in Equation \eqref{eqn:maxLBEst} as large maximization problem over
$\vec\zeta,\vec\theta_1,\dots,\vec\theta_n$. This problem can easily be solved
using tailored quasi-Newton methods. However, the new problem consists of a
maximin problem where we have to maximize over
$\vec\zeta,\vec\theta_1,\dots,\vec\theta_n$ but minimize over
$\vec\rho_1,\dots,\vec\rho_n$. We have to minimize over the latter because they
form a lower bound on $D_i(\vec\zeta)$ which is subtracted in
the lower bound on the log marginal likelihood.
Moreover, our approximation
is no longer guaranteed to be a lower bound.

We have created an implementation in which the optimal $\vec\rho_i$ is found
for each $(\vec\zeta, \vec\theta_i)$ each time the lower bound or its gradient
is evaluated. The optimization problem for $\vec\rho_i$ is easily solved but
still adds a substantial overhead to each lower bound evaluation, thus,
increasing the computation time substantially. While we do expect that the
maximin problem can be efficiently solved, we have opted for using
\ac{AGHQ} \citep{Crowther16} to approximate
$D_i(\vec\zeta)$. It is easier to approximate
$D_i(\vec\zeta)$ using \ac{AGHQ} compared to
$\bar L_i(\vec\zeta)$ in Equation \eqref{eqn:margLNoDelay} for several reasons.
The integral is possibly lower dimensional as we do not have to marginalize
over $W_{ie}$'s for recurrent events. In particular, we only have a $R$
dimensional integral if all terminal events do not have frailties. The
integrand for $D_i(\vec\zeta)$ is also easier to compute pointwise
as there are fewer factors than for $\bar L_i(\vec\zeta)$.


\section{Simulation studies and applications}\label{sec:simNApplication}

\begin{table}[tb!]
\centering
\caption{
Information about the simulation studies and applications in the paper.
$L$ is the number of longitudinal variables and the second column shows the number
of random effects per individual. $n$ is the number of individuals and the
subsequent four columns show the number of terminal event types,
the number of recurrent event types, the number of association
parameters and whether there is delayed entry, respectively.}
\label{tbl:tblExStats}
\begingroup\small
\makebox[\textwidth][c]{
\begin{tabular}{lrrrrrrr}
  \toprule
   Example
     & $L$
     & \# random effects
     & $n$
     & \# terminal
     & \# recurrent
     & \# associations
     & delayed
     \\
  \midrule
   Univariate simulation  &
     1 & 3 & 500 &  1 & 0 & 1/2 & No \\
   Multivariate simulation &
     2 & 7/5 & 1000 &  1 & 1 & 4 & No/yes \\
   pbc2 &
     3 & 6 & 154 &  1 & 0 & 3 & No\\
   Breast cancer \ &
     2 & 2/4 & \nBreastObs & 1 & 0 & 2 & Yes \\
   \bottomrule
\end{tabular}
}
\endgroup
\end{table}

Results are shown for simulation studies and applications in this section.
Table \ref{tbl:tblExStats} provides information about the examples
and shows that a broad range of models have been used.

\subsection{Simulations studies}\label{sec:Sims}

We start by presenting simulation studies that compare our \ac{GVA} with other
frequentist implementations.
The studies compare the computation time, bias,
and the inverse of the observed information matrices of the methods.
We also performed a coverage analysis of approximate profile likelihood based
confidence intervals from our \ac{GVA}.

We compared our method with joineRML \citep{Hickey18} and
the JM package.
joineRML is implemented for semi-parametric models
with a non-parametric baseline hazard.
The estimation method in joineRML is a \ac{MC} \ac{EM} algorithm.
The JM package can estimate models like ours with a parametric
but possibly flexible baseline hazard, but only
with one longitudinal variable and without frailty.
Thus, we used the same flexible B-spline with ten knots for the baseline hazard
in both our method and the JM package to get comparable estimates to those of
joineRML and between our method and the JM package.
The knots were placed at quantiles of the uncensored event times of
each data set.
Gauss-Hermite quadrature is used in the JM package to marginalize out the
random effects in an \ac{EM} algorithm and direct maximization of
the likelihood.
This can yield an arbitrary precise estimate of the maximum likelihood
estimator but scales poorly in the number of random effects per
individual.
Our method and the JM package provide almost identical models
except that JM uses the alternative parameterization described in
Appendix \ref{sec:alternativeParam}.
Thus, the two methods only give close to identical models as
$\vec z_{ih}(t)$ will be chosen to be more flexible than
$\vec x_{il}(t)$ but there is no linear map to $\vec x_{il}(t)$. We set the
non-likelihood-based convergence parameters, "tol1" and "tol2," to zero to
reduce the number of times
the maximum lower bound of the likelihood exceeded the achieved maximum
likelihood.

\begin{knitrout}\footnotesize
\definecolor{shadecolor}{rgb}{0.969, 0.969, 0.969}\color{fgcolor}\begin{figure}

{\centering \includegraphics[width=\maxwidth]{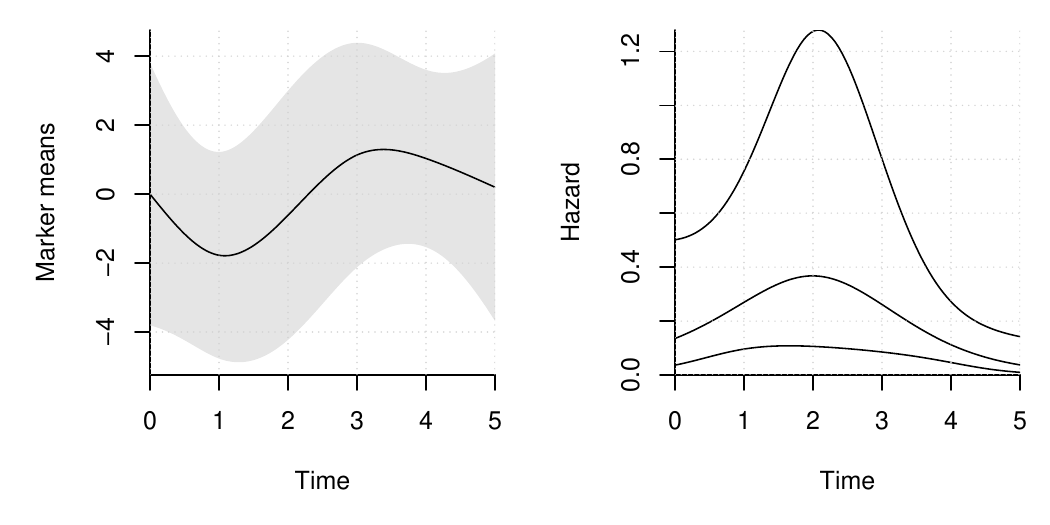}

}

\caption[The time-varying intercept of the longitudinal variables along with 95\% pointwise probability intervals for the individuals' mean curves (left plot), and  25\%, 50\%, and 75\% pointwise quantiles for the baseline hazards (right plot)]{The time-varying intercept of the longitudinal variables along with 95\% pointwise probability intervals for the individuals' mean curves (left plot), and  25\%, 50\%, and 75\% pointwise quantiles for the baseline hazards (right plot).}\label{fig:first_sim_setup}
\end{figure}

\end{knitrout}

\subsubsection{Univariate example}

We start by summarizing results from
the simulation studies we
ran with a univariate longitudinal variable and terminal time-to-event outcome.
We used a low degree spline for the baseline hazard.
The current value was used for the association,
with independent and random right-censoring times, and
administrative right-censoring.
The event was terminal and no longitudinal variables were observed
after the observed time (minimum of the censoring and event time).
The longitudinal variables had a random effect modeled with
a spline with an intercept and two additional degrees of freedom.
The longitudinal variables had a time-varying
intercept modeled with a spline using five degrees of freedom.
A fixed effect covariate was included in both the longitudinal variable sub-model
and the survival sub-model.
The gap times
between the observation times, $s_{ij}$'s, were drawn
from an exponential distribution with rate 1.5 and
the longitudinal variable was
observed at baseline.
Figure \ref{fig:first_sim_setup}
shows the longitudinal variables' mean curve and pointwise quantiles of the
hazard. A detailed description of the simulation study and the other
simulation studies are given in Appendix
\ref{sec:simComments}. All the simulations were run on a Laptop  with an
Intel\textsuperscript{\small\textregistered}~Core\texttrademark
~i7-8750H CPU and the code was compiled with GCC 10.1.

\begin{table}[tb!]
\centering
\caption{Bias estimates of model parameters (rows 1-5), \acs{MC} standard errors of the bias estimates (rows 6-10), and the root mean square errors (rows 11-15). Results from our \acs{GVA} are shown with 15 and 32 nodes used to compute the approximate expected cumulative hazard. The JM package is used with four and seven nodes used to marginalize out the random effects and 15 nodes used to compute the cumulative hazard. The last two columns show the average Euclidean norm of the error of the estimated covariance matrices and the average computation time in seconds.}
\label{tbl:firstSimBias}
\begingroup\small
\makebox[\textwidth][c]{
\begin{tabular}{lrrrrrrrr}
  \hline
Method & $\beta_{11}$ & $\beta_{12}$ & $\sigma^2_1$ & $\gamma_{11}$ & $\gamma_{12}$ & $\alpha_{11}$ & $\lVert\hat{\mat\Psi} - \mat\Psi\rVert$ & Time \\
  \hline
Bias &  &  &  &  &  &  &  &  \\
  GVA 15 & 0.00783 & -0.00438 & -0.00047 & -0.01112 & 0.00655 & -0.00013 & 2.16449 & 2.024 \\
  GVA 32 & 0.00783 & -0.00438 & -0.00047 & -0.01101 & 0.00658 & -0.00031 & 2.16464 & 2.675 \\
  joineRML & 0.01828 & -0.00434 & -0.00061 & -0.01241 & 0.00614 & 0.00395 & 2.19245 & 719.268 \\
  JM 4 & -0.00224 & -0.00410 & -0.00209 & -0.01187 & 0.00633 & 0.00229 & 2.18567 & 42.767 \\
  JM 7 & 0.00436 & -0.00424 & -0.00043 & -0.01013 & 0.00637 & -0.00575 & 2.17030 & 224.622 \\
  SE &  &  &  &  &  &  &  &  \\
  GVA 15 & 0.00840 & 0.00610 & 0.00105 & 0.01194 & 0.00814 & 0.00613 & 0.08308 & 0.008 \\
  GVA 32 & 0.00840 & 0.00610 & 0.00105 & 0.01193 & 0.00814 & 0.00613 & 0.08311 & 0.009 \\
  joineRML & 0.00832 & 0.00606 & 0.00105 & 0.01195 & 0.00802 & 0.00614 & 0.08658 & 30.405 \\
  JM 4 & 0.00845 & 0.00591 & 0.00105 & 0.01191 & 0.00790 & 0.00613 & 0.08698 & 0.363 \\
  JM 7 & 0.00838 & 0.00607 & 0.00105 & 0.01198 & 0.00812 & 0.00627 & 0.08505 & 1.694 \\
  RMSE &  &  &  &  &  &  &  &  \\
  GVA 15 & 0.08395 & 0.06085 & 0.01048 & 0.11932 & 0.08128 & 0.06099 &  &  \\
  GVA 32 & 0.08395 & 0.06085 & 0.01048 & 0.11921 & 0.08128 & 0.06098 &  &  \\
  joineRML & 0.08481 & 0.06046 & 0.01050 & 0.11959 & 0.08005 & 0.06118 &  &  \\
  JM 4 & 0.08409 & 0.05896 & 0.01064 & 0.11912 & 0.07882 & 0.06108 &  &  \\
  JM 7 & 0.08351 & 0.06050 & 0.01050 & 0.11959 & 0.08100 & 0.06265 &  &  \\
   \hline
\end{tabular}
}
\endgroup
\end{table}
\begin{table}[tb!]
\centering
\caption{Additional statistics for the \ac{GVA} model fits and model fits from the JM package (for the models included in Table \ref{tbl:firstSimBias}) along with \ac{MC} standard errors. The average maximum lower bound and maximum log likelihood are shown in the first column. The last column shows the average relative difference to the covariance matrix for the fixed effects and the association parameter from the JM package using seven nodes.}
\label{tb:firstSimLL}
\begingroup\small
\makebox[\textwidth][c]{
\begin{tabular}{lrr}
  \hline
Method & Log likelihood & Relative covariance error \\
  \hline
Estiamte &  &  \\
  GVA 15 & -3070.070 & 0.014307 \\
  GVA 32 & -3070.116 & 0.014382 \\
  JM 4 & -3070.311 & 0.122795 \\
  JM 7 & -3069.541 & 0.000000 \\
  SE &  &  \\
  GVA 15 & 7.590 & 0.001375 \\
  GVA 32 & 7.590 & 0.001376 \\
  JM 4 & 7.587 & 0.002540 \\
  JM 7 & 7.590 & 0.000000 \\
   \hline
\end{tabular}
}
\endgroup
\end{table}

We simulated 100 data sets, constrained by the performance of the slowest package compared, each comprising 500 individuals.
The average number of observed longitudinal variables
per individual
was 3.83 with a standard deviation of
2.98. On average,
42.4\%
of the time-to-event outcomes were censored.
Table \ref{tbl:firstSimBias} shows the bias estimates of the comparable
parameters along with root mean square errors and average computation
times. We show \ac{MC} standard errors of the estimates as these are relevant
for judging if the bias estimates are extreme and as one can get the actual standard
deviation the estimates by multiplying by the square root of the number of
samples.
All three methods gave comparable estimates with none showing
any evidence of bias. However, our \ac{GVA} was two orders of magnitude
faster than the JM package with seven Gauss-Hermite quadrature nodes
and the joineRML package.
We utilized four threads with our \ac{GVA}, whereas the other methods operate in a single-threaded mode. This setup demonstrates the efficiency of our package on a laptop. For comparison, the single-thread computation time is less than four times the reported multi-threaded time.
Moreover, we spent an average of 0.94 and 1.32 seconds computing the observed information matrix with our \ac{GVA} for 15 and 32 nodes, respectively. This computation time, included in Table \ref{tbl:firstSimBias}, can be significantly reduced, as outlined in Appendix \ref{sec:implementation}. We report it separately because it constitutes a substantial portion of the total computation time.
Table \ref{tb:firstSimLL} shows the average maximum lower bound and log
likelihood. The average difference between the two was small and
the maximum lower bound was always smaller as expected.
Moreover, we also compared the covariance matrix from the observed information
matrix from the JM package with seven nodes with the other methods.
Particularly, we considered the part of the covariance matrix for
$(\beta_{11},\beta_{12},\gamma_{11},\gamma_{12},\alpha_{11})$ as these are comparable
after accounting for the different parameterizations. We computed the Euclidean
norm of the difference between the covariance matrices and scaled the norm by
the Euclidean norm the covariance matrix part from the JM package with seven nodes.
As shown by Table \ref{tb:firstSimLL}, the
\ac{GVA} gave close to identical covariance matrices but
it was not sufficient to use four nodes with the JM package. The results also
show that it did not make a big difference to increase the number of
quadrature nodes used to approximate the integral
in Equation \eqref{eqn:expecHaz}.

\begin{table}[tb!]
\centering
\caption{Statistics for the \acs{GVA} model fits and model fits from the JM package. Models are the same as those used in Table \ref{tbl:firstSimBias} but with both current value and slope assocation. The first two columns show bias estimates, \acs{MC} standard errors of the estimates, and the root mean square errors for the association parameters for the current value, $\alpha_{111}$, and the slope, $\alpha_{112}$. The next three columns show the average Euclidean norm of the error of the estimated covariance matrices, the average maximum lower bound or log marginal likelihood, and the average computation time in seconds, respectively.}
\label{tbl:currentNSlope}
\begingroup\small
\makebox[\textwidth][c]{
\begin{tabular}{lrrrrr}
  \hline
Method & $\alpha_{111}$ & $\alpha_{112}$ & $\lVert\hat{\mat\Psi} - \mat\Psi\rVert$ & Log likelihood & Time \\
  \hline
Bias &  &  &  &  &  \\
  GVA 15 & -0.0029 & 0.0055 & 2.501 & -2927.7 & 2.18 \\
  GVA 32 & -0.0035 & 0.0064 & 2.502 & -2927.7 & 2.98 \\
  JM 4 & -0.0081 & 0.0244 & 2.507 & -2927.2 & 69.95 \\
  JM 7 & -0.0088 & 0.0233 & 2.516 & -2927.2 & 279.79 \\
  SE &  &  &  &  &  \\
  GVA 15 & 0.0057 & 0.0083 & 0.089 & 8.1 & 0.01 \\
  GVA 32 & 0.0057 & 0.0083 & 0.089 & 8.1 & 0.01 \\
  JM 4 & 0.0059 & 0.0089 & 0.089 & 8.1 & 0.62 \\
  JM 7 & 0.0059 & 0.0090 & 0.090 & 8.1 & 2.14 \\
  RMSE &  &  &  &  &  \\
  GVA 15 & 0.0568 & 0.0828 &  &  &  \\
  GVA 32 & 0.0569 & 0.0828 &  &  &  \\
  JM 4 & 0.0590 & 0.0915 &  &  &  \\
  JM 7 & 0.0595 & 0.0922 &  &  &  \\
   \hline
\end{tabular}
}
\endgroup
\end{table}

We also extended the first model to a current value and slope for the
association.
Table \ref{tbl:currentNSlope} shows the results.
The maximum lower bound was larger than the maximum log likelihood from the JM
package
$9$\%
of the time. We are not sure why.
It could be related to the different
parameterization although the baseline hazard was flexible with a
B-spline using ten knots. There were no evidence of bias with the
\ac{GVA} but some evidence of bias with the JM package with the association
parameter for the slope. The GVA was again much faster.

\begin{table}[tb!]
\centering
\caption{Statistics for the \ac{GVA} model fits for the simulations with the multivariate longitudinal variables and time-to-event outcomes. The first five columns show bias estimates, \acs{MC} standard errors of the estimates, root mean squared errors and coverage of Wald based 95\% confidence intervals for the association parameters and the scale parameter of the frailty. The last three columns show the average Euclidean norm of the error of the estimated covariance matrices, the average computation time in seconds and coverage of 95\% profile likelihood-based confidence intervals for the first association parameter, $\alpha_{11}$, respectively.}
\label{tbl:multBoth}
\begingroup\small
\makebox[\textwidth][c]{
\begin{tabular}{lrrrrrrrr}
  \hline
Metric & $\alpha_{11}$ & $\alpha_{12}$ & $\alpha_{21}$ & $\alpha_{22}$ & $\xi_2^2$ & $\lVert\hat{\mat\Psi} - \mat\Psi\rVert$ & Time & Coverage profile \\
  \hline
Bias & -0.0003 & 0.0006 & 0.0010 & 0.0001 & -0.0126 & 3.024 & 90.42 & 0.9390 \\
  SE & 0.0012 & 0.0012 & 0.0005 & 0.0005 & 0.0009 & 0.020 & 0.34 &  \\
  RMSE & 0.0385 & 0.0375 & 0.0162 & 0.0156 & 0.0315 &  &  &  \\
  Coverage & 0.9400 & 0.9550 & 0.9410 & 0.9590 & 0.9170 &  &  &  \\
   \hline
\end{tabular}
}
\endgroup
\end{table}

\begin{table}[tb!]
\centering
\caption{Statistics from a simulation study with delayed entry and with a model with a random intercept and slope for each of the two longitudinal variables.}
\label{tbl:multDelayedBoth}
\begingroup\small
\makebox[\textwidth][c]{
\begin{tabular}{lrrrrrrrr}
  \hline
Metric & $\alpha_{11}$ & $\alpha_{12}$ & $\alpha_{21}$ & $\alpha_{22}$ & $\xi_2^2$ & $\lVert\hat{\mat\Psi} - \mat\Psi\rVert$ & Time & Coverage \\
  \hline
Bias & 0.0010 & 0.0045 & -0.0007 & 0.0005 & -0.0172 & 0.080 & 170.51 & 0.9520 \\
  SE & 0.0029 & 0.0029 & 0.0014 & 0.0014 & 0.0011 & 0.001 & 0.25 &  \\
  RMSE & 0.0918 & 0.0917 & 0.0440 & 0.0456 & 0.0391 &  &  &  \\
  Coverage & 0.9510 & 0.9460 & 0.9540 & 0.9440 & 0.9220 &  &  &  \\
   \hline
\end{tabular}
}
\endgroup
\end{table}

\subsubsection{Multivariate example}

We performed a more complex simulation study with multivariate
longitudinal variables and time-to-event outcomes by extending the first
simulation study.
The fixed effects and random effects for the second longitudinal
variable were set similarly to the first one defined as in the previous
simulation study.
The random effects vector, $\vec U_i$, was six dimensional.
Both the random effects for the splines and the
error terms, $\vec\epsilon_{ij}$, were correlated.
The association parameters were set to have an effect like before.
The observation process was informative with
an opposite sign association with the markers, such that
individuals with a higher hazard of the terminal time-to-event had a lower
likelihood of having observations for the longitudinal variables.
A log-normal frailty was included for the
observation process leading to seven unobserved random effects for
each individual.
The baseline hazard for the observation process was from a Weibull model
and the sample size was increased to
1000 individuals
to illustrate that our method works on larger data sets.
See Appendix \ref{sec:simComments} for further details.

The optimization problem involved 35070 parameters as there
were
70 model parameters and 35 variational parameters
per
individual. Each individual had an average of
4.8
observed markers with
a standard deviation of 8.08.
The results of the simulation study with 1000 sampled data
sets are shown in Table \ref{tbl:multBoth}. There
was no evidence of bias for any of the parameters (including those not shown;
see the plots in Appendix \ref{sec:simComments})
with the exception of the scale parameter of the frailty. The latter is not surprising
given that the log-normal distribution may not be a good approximation of the
conditional distribution of the frailty. Moreover, it did not seem to affect
the other parameters in the model. It took an average of
33.9 seconds to compute
the information matrix
which is included in the estimation time. The coverage of a
95\% profile likelihood-based confidence interval for the
first association parameter, $\alpha_{11}$,
was close to that of the Wald-based and the nominal level.
It took an average of
151 seconds to compute the profile
likelihood-based confidence interval.

We added delayed entry to the model in a second multivariate simulation study.
The data set was simulated such that individuals and delayed entry times were
sampled until 1000 individuals were observed. All parts of the model
were as before with the exception of the random effects in the
longitudinal variable sub-models, which were changed to a
random intercept and random slope in time. Thus, there are four random effects
in the longitudinal variable sub-models.
We do not have to marginalize out the frailty in
the delayed entry terms as
frailties for recurrent events do not enter into
Equation \eqref{eqn:delayedTerm}.
Table \ref{tbl:multDelayedBoth} shows the results from
1000 sampled data sets. The result are similar to
those in Table \ref{tbl:multBoth} without the delayed entry.
It took an average of
113.8 seconds to compute
the observed information matrix and
126.8 seconds to compute the profile
likelihood-based confidence interval.
The large computation time despite the fewer random
effects is due to the \ac{AGHQ}.
The computational complexity of using $b$ quadrature nodes per
random effect and with $R$ random effects is $\bigO{b^R}$ which becomes an
issue when $R$ is larger.


\subsection{Applications}\label{sec:application}

\begin{table}[tb!]
\centering
\caption{Point estimates of the three marker model fitted by Hickey and others (2018) along with estimated standard errors from the semiparametric model from the joineRML package, our \acs{GVA}, and the GVA estimated with the limited-memory \acs{BFGS} algorithm. The last column shows the estimation time in seconds.}
\label{tbl:pbc2Fit}
\begingroup\small
\makebox[\textwidth][c]{
\begin{tabular}{lrrrrrrrrrrr}
  \hline
Method & $\beta_{11}$ & $\beta_{12}$ & $\beta_{21}$ & $\beta_{22}$ & $\beta_{31}$ & $\beta_{32}$ & $\gamma_{11}$ & $\alpha_{11}$ & $\alpha_{12}$ & $\alpha_{13}$ & Time \\
  \hline
Estimate &  &  &  &  &  &  &  &  &  &  &  \\
  joineRML & 0.555 & 0.201 & 3.554 & -0.125 & 0.830 & -0.058 & 0.046 & 0.816 & -1.701 & -2.232 & 82.484 \\
  GVA & 0.556 & 0.202 & 3.555 & -0.125 & 0.830 & -0.058 & 0.047 & 0.896 & -1.662 & -2.058 & 7.159 \\
  GVA BFGS & 0.556 & 0.202 & 3.555 & -0.125 & 0.830 & -0.058 & 0.047 & 0.896 & -1.662 & -2.058 & 66.121 \\
  SE &  &  &  &  &  &  &  &  &  &  &  \\
  joineRML  & 0.085 & 0.020 & 0.036 & 0.010 & 0.021 & 0.006 & 0.015 & 0.204 & 0.619 & 1.628 &  \\
  GVA  & 0.088 & 0.021 & 0.032 & 0.011 & 0.019 & 0.006 & 0.014 & 0.186 & 0.550 & 1.371 &  \\
   \hline
\end{tabular}
}
\endgroup
\end{table}

We fit a model with three longitudinal variables, $L = 3$,
to the pbc2
data set. \cite{Hickey18} have also used their joineRML package
to model these data using a
non-parametric baseline hazard. Their model included both a fixed and random
intercept and slope for each longitudinal variable
yielding six random effects per individual
and six fixed effect coefficients in the longitudinal variable sub-models
(a $\beta_{l1}$ for the intercept and a $\beta_{l2}$ for the slope). Thus, there
are three association parameters with one for each longitudinal variable. Age
is included in the survival sub-model with a conditional proportional hazard
effect. The
corresponding coefficient is denoted by $\gamma_{11}$. The equations for the
model are shown in Appendix \ref{sec:appliedComments}.

Table \ref{tbl:pbc2Fit} shows the estimates using joineRML and our \ac{GVA}.
The estimates differ slightly from those in
\cite{Hickey18} because of \ac{MC} errors as the model is refitted.
It took 7.16 seconds to estimate the
\ac{GVA} of which 2.19 seconds
were spent on computing the observed information matrix.
The relatively quicker estimation from the joineRML package compared to Section
\ref{sec:Sims}, is likely due to the few individuals (there was only
154 individuals).
We also estimated the \ac{GVA} with the limited-memory
\ac{BFGS} algorithm implementation from
the lbfgsb3c package \citep{Fidler20}. The results are essentially identical, as
expected, but the estimation time was much slower. This shows the
importance of using the quasi-Newton method.

\begin{knitrout}\footnotesize
\definecolor{shadecolor}{rgb}{0.969, 0.969, 0.969}\color{fgcolor}\begin{figure}

{\centering \includegraphics[width=\maxwidth]{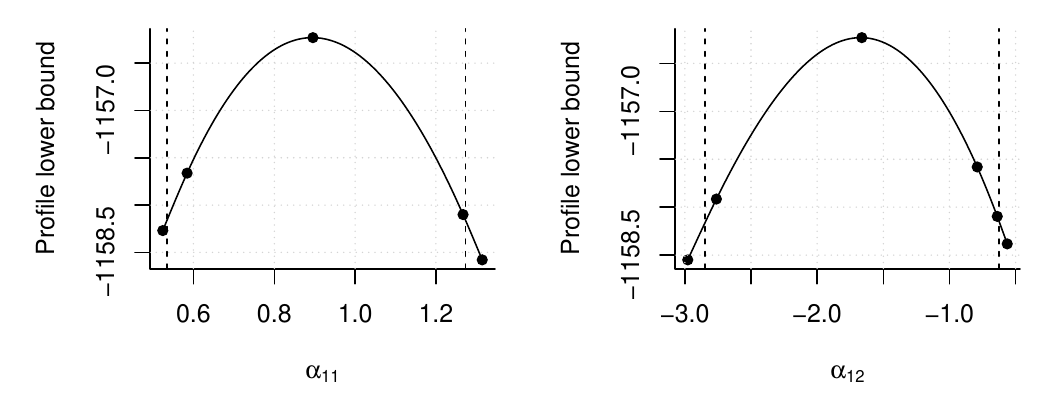}

}

\caption[Profile lower bound plots for two of the association parameters in the pbc2 model estimated with the \ac{GVA}]{Profile lower bound plots for two of the association parameters in the pbc2 model estimated with the \ac{GVA}. The dashed vertical lines show the limits of the 95\% confidence intervals.}\label{fig:pbc2_show_pls}
\end{figure}

\end{knitrout}

The standard errors of the packages differed, with those from the \ac{GVA}
being smaller. This can possibly be explained by, in combination, the
non-parametric baseline hazard compared with the parametric,
the different point estimates, possible
\ac{MC} errors in the joineRML package, and
inaccuracies from using the \ac{VA}.
We computed approximate profile likelihood-based confidence intervals
for the association parameters. It took
60.32 seconds to compute the
intervals. The 95\% confidence intervals were
$(0.53, 1.27)$,
$(-2.85, -0.62)$, and
$(-4.87, 0.61)$
for $\alpha_{11}$, $\alpha_{12}$, and $\alpha_{13}$, respectively.
This is not far from the Wald type confidence intervals
as can be seen from the standard errors and estimates in Table
\ref{tbl:pbc2Fit}.
This is also
clear from the profile lower bound plots shown in
Figure \ref{fig:pbc2_show_pls} which are close to quadratic functions.

\subsubsection{Mammography markers of breast cancer risk}

Mammography density (the amount of fibroglandular tissue, which is
radiographically dense and appears white on a mammogram) is an established
risk factor for breast cancer. The role of fatty tissue is less well understood
\citep{Shepherd12}. To illustrate our approach on a large data set
we studied these
associations using information from the Karma project \citep{Gabrielson17}.
The analysis presented here aligns with that of \citep{Illipse24}, where the
associated package is utilized, though the method itself is not detailed.
Longitudinal measurements on dense and fatty
tissue areas from screening visits, from
study entry to end of follow-up were included.
The time-to-event outcome was time to
the first breast cancer diagnosis. Right-censoring was due to death or
administrative right-censoring at the end of 2018.
Attained age was used as the time scale and there was
left truncated data -- women were followed from time to study entry,
at a screen as part of the mammography screening program
(which starts at age 40) and all women had no prior diagnosis of breast cancer.
In our analysis we included \nBreastObs~women and
188037 longitudinal measurements.

\begin{table}[tb!]
\centering
\caption{
Point estimates for the association parameters and
variances of the random effects for joint models for breast cancer,
dense tissue and fatty tissue, with estimated standard errors.
$\alpha_{11}$ and $\alpha_{12}$ are the
association parameters for the dense and fatty tissue, respectively.
$\psi_1^2$ and $\psi_2^2$ are the variances of the random intercept and slope,
respectively, for the dense tissue. $\psi_3^2$ and $\psi_4^2$ are defined
analogously for the fatty tissue.
The last two
columns show the maximum lower bound and the computation time in hours,
respectively.}
\label{tbl:breastCancer}
\begingroup\small
\makebox[\textwidth][c]{
\begin{tabular}{lrrrrrrrr}
  \toprule
     Model
     & $\alpha_{11}$
     & $\alpha_{12}$
     & $\psi_1^2$
     & $\psi_2^2$
     & $\psi_3^2$
     & $\psi_4^2$
     & Lower bound
     & Time
     \\
  \midrule
   Estimate \\
   Without slopes
     & 0.1316 & 0.03045 & 3.777 & & 3.795 & & -587141.9 & 1.174 \\
   With slopes
     & 0.1436 & 0.02663 & 3.675 & 0.002327 & 3.421 & 0.002281 & -586179.6 & 38.490 \\
   SE \\
   Without slopes
     & 0.0128 & 0.01600 & 0.023 & & 0.023 & & & \\
   With slopes
     & 0.0133 & 0.01711 & 0.052 & 0.000161 & 0.052 & 0.000174 & & \\
   \bottomrule
\end{tabular}
}
\endgroup
\end{table}

We estimated two models: one with random intercepts for both dense and fatty tissue, and another with random slopes for time in addition to the intercepts. More complex random effect models were considered but could not be fit due to the limited observations per individual.
Moreover, we modeled the
square root of the fatty and dense tissue because of a skewed marginal
distribution.
All three sub-models included BMI which was only observed at baseline.
The equations for the
models are shown in Appendix \ref{sec:appliedComments}.
The model was estimated on a server with an
Intel\textsuperscript{\small\textregistered}~Xeon\textsuperscript{\small\textregistered}
CPU E5-2660 v2 using six cores for the model without the random slopes and
eight cores for the model with the random slopes.

The estimated association parameters and random effects' covariances
are shown in Table \ref{tbl:breastCancer} along with maximum lower bounds and
the computation times.
The estimated association parameter for the dense tissue, $\alpha_{11}$,
was highly significant in both cases and
shows that a one standard deviation change in the
women-specific intercept yields a factor
$\exp(\alpha_{11}\psi_1) = 1.291$ higher
hazard of breast cancer in the model without the random slopes. The association
parameter for the fatty tissue shows a non-significant association in both
models. The change in the lower bound from adding the random slopes was relatively
small but in absolute terms large. This is also reflected in the quite small
but significant
variances of the random slopes, $\psi_2^2$ and $\psi_4^2$. As expected
there was a negative correlation between the two random intercepts and the
two random slopes showing an inverse relationship between the dense
and fatty tissue given BMI.

The model without and with the random slopes had 300446 and 841182 parameters
in total, respectively.
The longer computation times compared with the simulations
accounting for the sample size
are attributable to three factors:
the delayed entry which terms are approximated with \ac{AGHQ},
the CPU is slower in benchmarks compared with the one used in the simulations
and more iterations were used to reach convergence.


\section{Discussion}\label{sec:discussion}
Estimation of joint models with multiple longitudinal markers and survival
outcomes is increasingly in demand with today's health and population registers.
The scale of the data sets makes it important to have fast and
scalable methods.
We have shown that a \ac{GVA} can be implemented to provide very fast
estimates for a flexible class of joint models.
In particular, fast estimation is achievable by exploiting that the
lower bound from the \ac{GVA} is partially separable. We also showed that
the lower bound for a large class of \acp{VA} only requires computation of the
moment generating function, the first two moments and the entropy of the
variational distribution. This makes it
feasible to use other distributions such
as the multivariate skew-normal distribution \citep{ormerod11}.
Nevertheless, our simulation studies and applied examples show that the bias of
the \ac{GVA} is negligible or non-existent, except for the scale parameters
of the log-normal frailties.

Our work and implementation can be extended in many ways. The method is
easily extended to support both left-censoring and interval-censoring. This
only requires additional computation of the approximate expected cumulative
hazard. The optimization method can be changed from an inexact
quasi-Newton method to an inexact Newton method \citep{nocedal06} by
replacing the \ac{BFGS} approximations of each lower bound
term's Hessian with the Hessian.
Moreover,
the fixed effects in the longitudinal variable sub-models and the survival
sub-models along with the means from the \ac{GVA}
can efficiently be profiled out.
These parameters can be estimated with the Newton-Raphson method given the
covariance matrix of the random effects and the covariance matrices from the
\ac{GVA}. Both additions may reduce the computation time.

We only considered continuous longitudinal variables which are conditionally
normally distributed.
Our approach could be extended to \acp{GLMM}.
This would only change the lower bound terms from the
conditional density of longitudinal variables.
Moreover, \cite{Ormerod12} show that these terms, at worst, involve many
one-dimensional integrals which can be solved quickly with
numerical integration for many combinations of link functions and distributions.
The quasi-Newton method we have developed for partially separable problems
can still be used and our work with \acp{GLMM} suggest
that the lower bound can be optimized very quickly.

\section*{Acknowledgments}

This work was partially supported by the Swedish e-Science Research Center and
the Swedish Research Council (grants 2019-00227 and 2020-01302).
Alessandro Gasparini was supported by a Swedish Cancer Society grant
(2020-0714).

\bibliographystyle{biom}
\bibliography{VAJointSurv-paper}

\section*{Supporting Information}
An R package containing the implementation is available at
https://github.com/boennecd/VA\-Joint\-Surv and on CRAN.

\appendix

\section{Alternative parameterization of the hazard}\label{sec:alternativeParam}

An alternative parameterization, which is also commonly used, is %
\begin{equation}\label{eqn:meanHazRelation}
h_{ie}(t\mid \vec u, w) = \exp\left(
  \vec z_{ie}(t)^\top\vec \gamma_e^* +
  \vec\alpha_e^\top\vec\mu_i(t, \vec u) + w_{ie}
  \right).
\end{equation}
The parameterization is used for instance in the JM package \citep{Rizopoulos10}
which we compare.
For this parameterization, the hazards are linked to the longitudinal variables'
latent mean (or a transformation
thereof) instead of the deviation from the population.
We discuss this parameterization as it is common and we want to motivate why we
choose the alternative parameterization in Equation \eqref{eqn:subhaz}.

Importantly, the interpretation of the association parameters will be the same
in both parameterizations and the model in Equation \eqref{eqn:meanHazRelation}
and \eqref{eqn:subhaz}
are equivalent in some cases.
This is true if there is a linear map from $\vec z_{ie}(t)$ to the fixed effect
design vectors in the longitudinal variable sub-models.
For example, consider $L = 1$ and assume that there is an invertible matrix
$\mat G = (\mat G_1, \mat G_2)$ such that %
$$
\vec z_{ie}(t) =
  \mat G_1 \bar{\vec z}_{ie}(t) + \mat G_2\vec x_{i1}(t)
$$%
for some $\bar{\vec z}_{ie}(t)$. If we substitute for
$\vec z_{ie}(t)$ and $\mu_{i1}(t,u)$ into Equation~\eqref{eqn:meanHazRelation},
then%
\begin{align*}
h_{ie}(t\mid \vec u_i, w_{ie}) &= \exp\left(
  \vec z_{ie}(t)^\top\vec \gamma_e^* +
  \alpha_{e1}\mu_{i1}(t, \vec u_i) + w_{ie}
  \right) \\
&= \exp\Big(
  \bar{\vec z}_{ie}(t)^\top\mat G_1^\top\vec\gamma_h^* +
  \vec x_{i1}(t)^\top\mat G_2^\top
  (\vec\gamma_e^* + \mat G_2^{+\top}\alpha_{e1}\vec\beta_1)  \\
&\hspace{45pt}+
  \alpha_{e1}\vec m_{i1}(t)^\top\vec u_i + w_{ie}
  \Big).
\end{align*}%
where $\mat G_2^+$ is the Moore–Penrose inverse of $\mat G_2$.
Thus, the fixed effects in the survival sub-model in the two parameterizations
can be identified with the identity
$\vec\gamma_e = \mat G^{-\top}(\vec\gamma_e^{*^\top}\mat G_1,
  (\vec\gamma_e^* + \mat G_2^{+\top}\alpha_{e1}\vec\beta_1)^\top\mat G_2)^\top$,
$\alpha_{e1}$ is the same and the two models are equivalent.

However, such a map may not exist,
particularly if $\vec z_{ie}(t)$ is chosen to be
more parsimonious than $\vec x_{il}(t)$. As the baseline hazard is often
chosen to be flexible in $\vec z_{ie}(t)$, the main difference is
either an omitted covariate or a time-varying covariate effect. Such omitted
effects on the hazard in the survival sub-model will
then be determined by the $\vec\alpha_e$s and $\vec\beta_l$s.
This will give an efficiency gain if the model assumption is true but it may be
difficult to justify in practice.
For example, if the markers include sex as a covariate, but the hazard does not,
then the only effect of sex on the hazard will be posited to be through the mean
marker values.
Moreover, the hazard in Equation
\eqref{eqn:subhaz} has computational advantages
in the quadrature rule used to compute the cumulative hazard as one does not
have to evaluate $\mat X_i(t)$ at each quadrature node.
For these reasons, we prefer the
parameterization in Equation~\eqref{eqn:subhaz}.

\section{Optimizing the lower bound}\label{sec:psqnNotes}
The lower bound in Equation \eqref{eqn:maxLBEst} is a joint optimization
problem over $\vec\zeta$ and $\vec\theta$, where
$\vec\theta = (\vec\theta_1^\top, \dots, \vec\theta_n^\top)^\top$.
It is partially
separable where each lower bound term, $\tilde L_i$, is a so-called element
function. For these problems, quasi-Newton methods can
be implemented to quickly find the maximum of the optimization problem
\citep{nocedal06}.

To explain this further, let $\vec\zeta\in\mathbb R^{v_1}$ and
$\vec\theta_1,\dots,\vec\theta_n\in\mathbb R^{v_2}$. The optimization
problem is to find the maximum of %
\begin{equation}\label{eqn:fullLB}
\tilde{L}(\vec\theta,\vec\zeta) =
  \sum_{i = 1}^{n}\tilde{L}_i(\vec\theta_i,\vec\zeta)
\end{equation}%
as a function of the $v_1 + nv_2$ parameters.
The problem quickly becomes high dimensional with many individuals. This implies
that \ac{BFGS} algorithm is not feasible because of storage of a
$(v_1 + nv_2)\times (v_1 + nv_2)$ matrix and subsequent matrix-vector products
with a $\bigO{(v_1 + nv_2)^2}$ computational complexity.
Limited-memory \ac{BFGS} can be used but
it is slow in practice as we show in Section
\ref{sec:application}. However, the Hessian of Equation \eqref{eqn:fullLB} is
very sparse. It consists of a dense $v_1\times v_1$ part from
$\nabla_{\vec\zeta\vec\zeta}^2\tilde{L}$,
a dense $v_1\times nv_2$ block from
$\nabla_{\vec\zeta\vec\theta}^2\tilde{L}$, and
a block diagonal matrix with $n$ blocks of size $v_2\times v_2$ from
$\nabla_{\vec\theta\vec\theta}^2\tilde{L}$. That is, an arrowhead matrix-like
structure. The idea then is to develop a
quasi-Newton method which preserves the sparsity of the Hessian.

Following \cite{nocedal06}, we
make $n$ \ac{BFGS} approximations
of the Hessian of each $\tilde{L}_i$. The $n$ approximations are
combined to
an approximation of the entire Hessian of $\tilde{L}$ which is
sparse unlike with the
\ac{BFGS} algorithm or its limited-memory version. The conjugate gradient
method can then be used to quickly and approximately
solve the linear equations in the resulting quasi-Newton method.
This only requires an implementation of the matrix-vector product between a
$v_1 + nv_2$ dimensional vector and the approximate Hessian.
We use a block diagonal preconditioning method in the conjugate gradient
method which is formed by ignoring the
$\nabla_{\vec\zeta\vec\theta}^2\tilde{L}$ block of the Hessian. This seems to
work well.

We have implemented the above in our psqn package with a
header-only \Cpp ~library. The entire algorithm is easy to compute in
parallel.  The  main computational bottlenecks are to compute
the lower bound terms, their gradient, updating the \ac{BFGS}
approximations, setup and apply the preconditioning, and computing the
matrix-vector product in the conjugate gradient method. All of these tasks
are easy to do in parallel.

\section{Implementation}\label{sec:implementation}
We have implemented the \ac{GVA} for the joint models. Our implementation
supports up to $L = 31$ types of longitudinal variables,
and an arbitrary number of types of time-to-event
outcomes. We support both terminal events, recurrent events and delayed entry.
Our implementation is written in \Cpp ~with support for
computation in parallel using
OpenMP. We use our quasi-Newton method for partially separable
problems in the header-only \Cpp ~library from our psqn package.
See Appendix \ref{sec:psqnNotes} for details. The psqn package
also provides an approximation of the Hessian given a user-defined gradient
which we need for the observation matrix in Equation \eqref{eqn:obsMat}. The
Hessian from psqn is based on numerical differentiation with Richardson
extrapolation \citep{Richardson27}
to refine the estimates. The software exploits that the Hessian of all the
model parameters can be computed as a sum of the Hessian of low dimensional
functions. However, the numerical differentiation is much
slower than an efficient
manual Hessian implementation which should be kept in mind when
seeing the computation times for the Hessian.

Starting values are found by fitting the time-to-event and
longitudinal variable sub-models
separately. The longitudinal variable sub-models are estimated for
each longitudinal variable
with the lme4 package \citep{Bates15}.
The starting values for the variational parameters are then
set using the best linear predictors of the random effects from the lme4 package
and the estimated unconditional random effect covariance matrix, $\mat\Psi$.
We use an extension of the reverse mode automatic differentiation library
developed by \cite{savine2018modern}. Our extension includes the matrix and
vector functions we need. The library is similar to the Adept library
\citep{Hogan14}. It is a fast implementation with eager evaluation of
the partial derivatives, and
makes use of expression templates to reduce the number of stored
partial derivatives and the overhead.
Nevertheless, an efficient manual
implementation of the gradient would be faster. However, our benchmarks show
that the time to compute the gradient is only three to five times greater than
that of evaluating the lower bound. Thus, the reduction in computation time
may not be large.

\section{Details for the simulation studies}\label{sec:simComments}

In this section, we provide some further details about the simulation studies
in Section \ref{sec:Sims}. The longitudinal variable in the first simulation
study was sampled from %
\begin{align*}
Y_{ij1} &= \beta_{11} + \tilde X_{i1}\beta_{12}
  + \vec\beta_{1,-(1:2)}^\top\vec g(s_{ij})
  + \vec m(s_{ij})^\top\vec U_{i1} + \epsilon_{ij1} \\
\epsilon_{ij1} &\sim N(0, \sigma_1^2) \\
\tilde X_{i1} &\sim N(0, 1) \\
\vec U_{i1} &\sim N^{(3)}(\vec 0, \mat\Psi)
\end{align*}%
where $\vec g$ is a natural cubic spline basis with knots at
time $0,1,\dots,5$,
$\vec\beta_{1,-(k:l)}$ is all the elements of $\vec\beta_1$ except those from index
$k$ to $l$, and $\vec m$ is a natural cubic spline basis
with an intercept and knots at times zero, two, and five.
The fixed effects' coefficients were set to
$\beta_{11} = 1$ and $\beta_{12} = -0.5$ and the
variance of the error term was $\sigma_1^2 = 0.25$.
The covariance matrix for the random effects was %
$$
\mat\Psi = \begin{pmatrix}
  9.2 & 1.92 & -0.96 \\
  1.92 & 4 & -2 \\
  -0.96 & -2 & 7.6
  \end{pmatrix}.
$$%

The survival sub-model for the terminal time-to-event outcome had a
conditional hazard of the form %
\begin{align*}
h_{i1}(t\mid\vec u) &= \exp\left(
  \tilde Z_{i1} \gamma_{11} + \vec\gamma_{1,-1}^\top\vec b(t) +
  \alpha_{11}\vec m(t)^\top\vec u
  \right) \\ %
\tilde Z_{i1} & \sim \text{Unif}(-1, 1)
\end{align*}%
where $\text{Unif}(a, b)$ denotes the uniform distribution on the interval
$(a, b)$ and $\vec b$ is a natural cubic spline basis with an intercept and knots at
times zero, two, and five. We set $\gamma_{11} = 0.4$ and
$\alpha_{11} = -1$.
An independent and random right-censoring were drawn from an
exponential distribution with rate 0.1667, and there was
administrative right-censoring at time 5.
Samples were drawn
from the conditional hazards by inverting the cumulative density function which
was approximated with Gauss–Legendre quadrature using 100 nodes. The error is
tiny for the latter approximation.

The model we estimated was %
$$
\hat h_{i1}(t\mid\vec u) = \exp\left(
  \tilde Z_{i1} \gamma_{11} + \tilde X_{i1}\gamma_{12} +
  \vec\gamma_{1,-(1:2)}^\top\tilde{\vec b}(t) +
  \alpha_{11}\vec m(t)^\top\vec u
  \right)
$$%
with $\tilde{\vec b}$ being more flexible than $\vec b$ as explained in
Section \ref{sec:Sims} to make the results comparable to the joineRML package.
Moreover, the additional
fixed effect in the hazard is needed with the JM package because of the
different parameterizations.

The second simulation study is very similar to the first but with the
conditional hazard given by %
\begin{align*}
h_{i1}(t\mid\vec u) &= \exp\left(
  \tilde Z_{1i} \gamma_{11} + \vec\gamma_{1,-1}^\top\vec b(t) +
  \alpha_{111}\vec m(t)^\top\vec u +
  \alpha_{112}\vec m'(t)^\top\vec u
  \right) \\
\vec m'(t) &= \left.\frac{\partial\vec m(s)}{\partial s}\right|_{s = t}.
\end{align*}%
That is, the hazard depends both on the current value and slope. We set
$\alpha_{111} = -1$ and
$\alpha_{112} = 0.5$.

The additional longitudinal variable in the last simulation study was
simulated from %
\begin{align*}
Y_{ij2} &= \beta_{21} + \tilde X_{i2}\beta_{22}
  + \vec\beta_{2,-(1:2)}^\top\vec g(s_{ij})
  + \vec m(s_{ij})^\top\vec U_{i2} + \epsilon_{ij2} \\
\begin{pmatrix} \epsilon_{ij1} \\ \epsilon_{ij2} \end{pmatrix} &\sim
  \begin{pmatrix}
    \sigma_1^2 & 0.5\sigma_1\sigma_2 \\
    0.5\sigma_1\sigma_2 & \sigma_2^2
  \end{pmatrix} \\
\tilde X_{i2} &\sim N(0, 1)
\end{align*} %
with $\vec\beta_2 = -\vec\beta_1$ and $\sigma_1 = \sigma_2$. The covariance
matrix for the random effects,
$\vec U_i = (\vec U_{i1}^\top, \vec U_{i2}^\top)^\top$, is set such that the
intercept of the two random splines are correlated with a correlation
coefficient of 0.5. This is achieved by setting %
%
%
$$
\mat \Psi = \begin{pmatrix}
    9.2 & 1.92 & -0.96 & 0.0815 & 0.4152 & 0.0501 \\
    1.92 & 4 & -2 & 0.4152 & 2.1147 & 0.2554 \\
    -0.96 & -2 & 7.6 & 0.0501 & 0.2554 & 0.0308 \\
    0.0815 & 0.4152 & 0.0501 & 9.2 & 1.92 & -0.96 \\
    0.4152 & 2.1147 & 0.2554 & 1.92 & 4 & -2 \\
    0.0501 & 0.2554 & 0.0308 & -0.96 & -2 & 7.6
  \end{pmatrix}.
$$

The conditional hazard for the terminal time-to-event was only altered by
adding the association with the additional longitudinal variable %
$$
h_{i1}(t\mid\vec U_i) = \exp\Big(
  \tilde Z_{i1} \gamma_{11} +
  \vec\gamma_{1,-1}^\top\tilde{\vec b}(t) \\
  + \alpha_{11}\vec m(t)^\top\vec U_{i1} +
  \alpha_{12}\vec m(t)^\top\vec U_{i2}
  \Big)$$%
with $\alpha_{11} = \alpha_{12} = 0.5$. Moreover, we include the additional
covariate from the second longitudinal variable, $\tilde X_{i2}$, when we
estimate the joint model.

The conditional hazard for the recurrent process for the observation process is
given by %
\begin{align*}
h_{i1}(t\mid\vec U_i, w_{i2}) &= \exp\left(
    \gamma_{21} + \gamma_{22}\log t  +
    \alpha_{21}\vec m(t)^\top\vec U_{i1} +
    \alpha_{22}\vec m(t)^\top\vec U_{i2} + w_{i2}
  \right). \\
W_{i2} & \sim N(0, \xi_2^2)
\end{align*}%
Thus, we have a Weibull baseline hazard with $\gamma_{22} = -0.2$ and
$\gamma_{21} = -0.5$ and we used $\alpha_{21} = \alpha_{22} = -0.5$.
The log-normal distributed frailty had $\xi_2^2 = 0.25$.
We made an expansion on log time
rather than time in the B-spline for the baseline hazard because of the
Weibull baseline hazard.
15 quadrature nodes to compute the approximate expected cumulative hazard in
Equation \eqref{eqn:expecHaz}.

The delayed entry time for each individual in the final simulation study was
sampled from a uniform distribution on zero to 0.5.
This yields a moderate chance that a possible individual is not observed with a
marginal survival probability at time 0.5 of
0.905 and a 95\% probability interval of
$(0.735, 0.979)$.
The covariance matrix for the random
effects was set to %
$$
\mat\Psi = \begin{pmatrix}
  0.5625 & -0.099 & 0.2812 & 0 \\
  -0.099 & 0.1089 & 0 & 0 \\
  0.2812 & 0 & 0.5625 & -0.099 \\
  0 & 0 & -0.099 & 0.1089
\end{pmatrix}
$$%
where the first and third dimension are for the random intercept and the
second and fourth dimension are for the random slopes.
The expectation in
Equation \eqref{eqn:delayedTerm} from the delayed entry was computed with
\ac{AGHQ} using 5 nodes for each of the four dimensions.

Figure \ref{fig:biasPlot} shows the estimated bias estimates divided by the
\ac{MC} standard errors. The parameters for the baseline hazards are not
included as the knots' position differ between fits as these are
determined by the observed event times.
The plot shows that the bias estimates are not extreme
except for the variance of the log normal frailty.
Figure \ref{fig:coveragePlot} shows a similar plot for the coverage of 95\%
Wald type confidence intervals. The conclusion is similar for the coverage
of the Wald type confidence intervals.

\begin{knitrout}\footnotesize
\definecolor{shadecolor}{rgb}{0.969, 0.969, 0.969}\color{fgcolor}\begin{figure}

{\centering \includegraphics[width=\maxwidth]{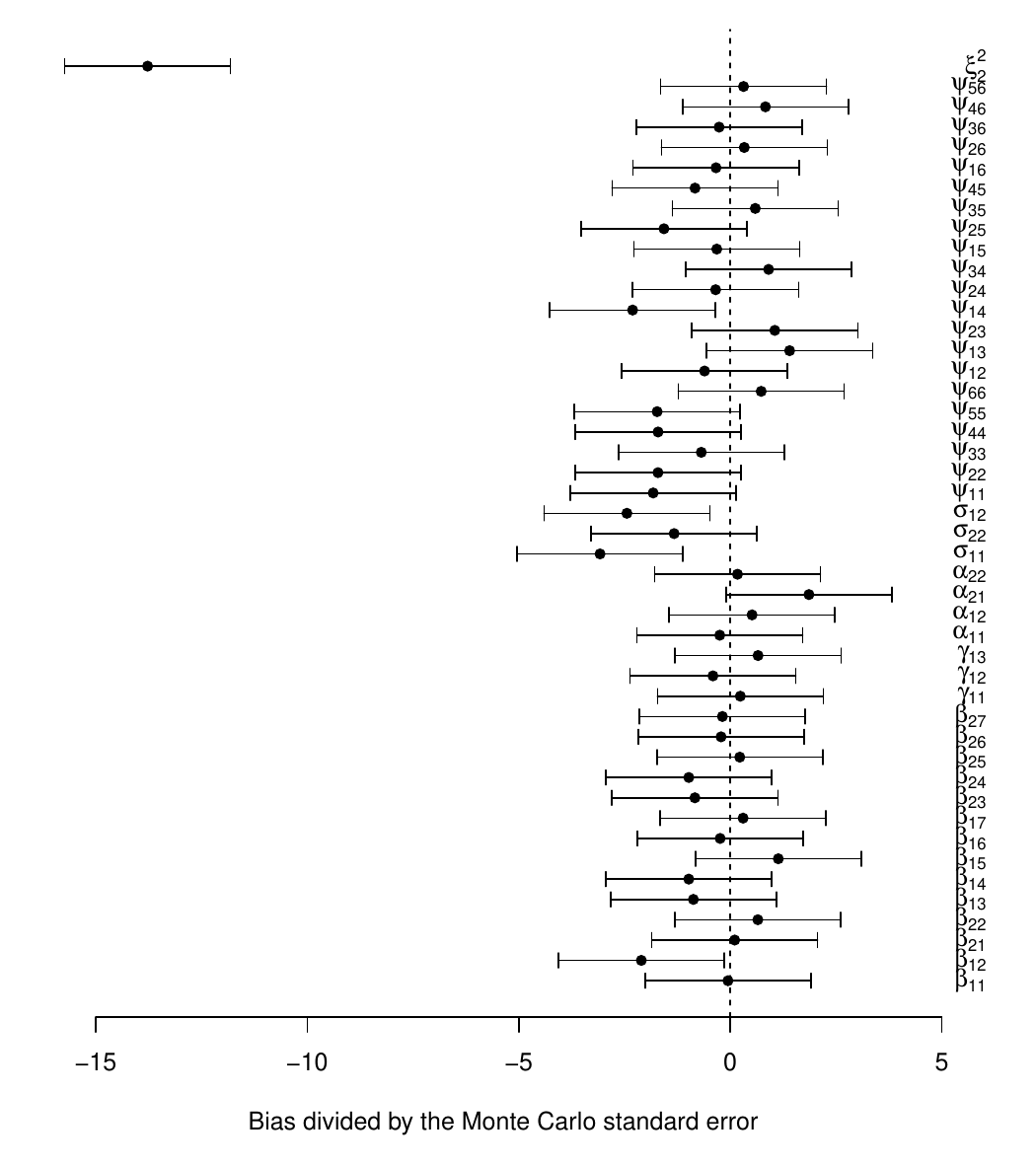}

}

\caption{Estimated bias divided by the \acs{MC} standard errors from the first multivariate simulation study which results are shown in Table \ref{tbl:multBoth}. The plot includes all the parameters in the model except for the baseline hazard parameters. The points are the scaled bias estimates and the lines show pointwise 95\% confidence intervals.}\label{fig:biasPlot}
\end{figure}

\end{knitrout}

\begin{knitrout}\footnotesize
\definecolor{shadecolor}{rgb}{0.969, 0.969, 0.969}\color{fgcolor}\begin{figure}

{\centering \includegraphics[width=\maxwidth]{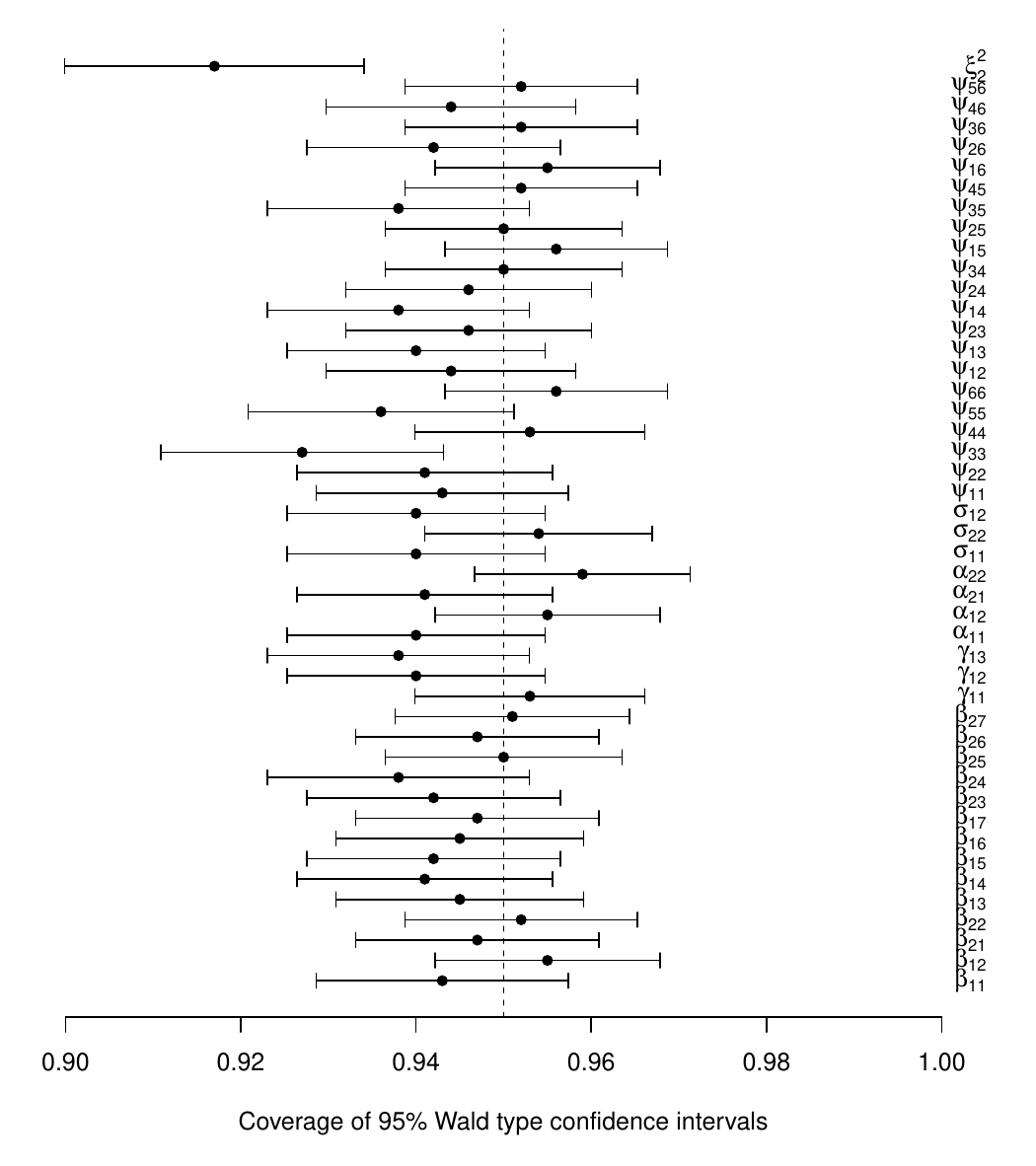}

}

\caption{Coverage of 95\% Wald type confidence intervals from the first multivariate simulation study which results are shown in Table \ref{tbl:multBoth}. The plot includes all the parameters in the model  except for the baseline hazard parameters. The points are the coverage estimates and the lines show pointwise 95\% confidence intervals for the coverage estimates.}\label{fig:coveragePlot}
\end{figure}

\end{knitrout}

\section{Details for the applied examples}\label{sec:appliedComments}

The model for the pbc2 data set was%
\begin{align*}
Y_{ijl} &=
  \beta_{l1} + U_{l1} + (\beta_{l2} + U_{l2}) s_{ij} + \epsilon_{ijl} \\
h_{i1}(t \mid \vec u_i) &=
  \exp\left(
    (\tilde z_i, \vec b(t)^\top)\vec\gamma_1
      + \alpha_{11}(U_{11} + t U_{12})
      + \alpha_{12}(U_{21} + t U_{22})
      + \alpha_{13}(U_{31} + t U_{32})
  \right)
\end{align*}%
where $\vec b$ is a vector of spline bases, $\tilde z_i$ is the age of
individual $i$ at baseline and for $l =1,\dots,3$.
We restricted the covariance $\mat\Sigma$ to be a diagonal matrix like in the
model estimated by \cite{Hickey18} to get comparable estimates.

The model for breast cancer, fatty and dense tissue was %
\begin{align*}
Y_{ijl} &=
  (\vec g(s_{ij})^\top, \tilde x_i)\vec\beta_l +
  U_{l1} + U_{l2} s_{ij} + \epsilon_{ijl} \\
h_{i1}(t \mid \vec u_i) &=
  \exp\left(
    (\tilde x_i, \vec b(t)^\top)\vec\gamma_1
      + \alpha_{11}(U_{11} + t U_{12})
      + \alpha_{12}(U_{21} + t U_{22})
  \right)
\end{align*}%
where $\tilde x_i$ is the BMI at study entry, $\vec g$ is
is a vector of spline bases using ten knots based on the observation times,
$\vec b$ is a vector of spline bases using ten knots based on the observed
event times
and for $l = 1,2$. A restricted model was also estimated where
$U_{12}$ and $U_{22}$ were omitted (a random intercept only model).

\end{document}